\documentclass[twocolumn, secnumarabic, amssymb, aps, prl, reprint,amsmath, superscriptaddress]{revtex4-2}
\usepackage{graphicx}

\newcommand{\schro}{Schr\"odinger~}
\usepackage{xcolor}
\usepackage{siunitx}
\usepackage{marvosym}

\newcommand{\todo}[1]{{\color{cyan} {#1}}}

\newcommand{\lettersection}[1]{\emph{\todo{#1}}.---}

\begin{document}

\title{Observation of Rydberg excitons in monolayer MoS$_2$ at room temperature by Imbert-Fedorov shift spectroscopy}%


\author{Xiaofeng Li}%
\affiliation{The Key Laboratory of Weak Light Nonlinear Photonics (Ministry of Education), School of Physics and Teda Applied Physics Institute, Nankai University, Tianjin 300071, China}
\affiliation{Department of Physics, City University of Hong Kong, Kowloon, Hong Kong SAR, China}
\author{Jiaxing Tu}
\affiliation{The Key Laboratory of Weak Light Nonlinear Photonics (Ministry of Education), School of Physics and Teda Applied Physics Institute, Nankai University, Tianjin 300071, China}
\author{Zhanyunxin Du}
\affiliation{Department of Physics, City University of Hong Kong, Kowloon, Hong Kong SAR, China}
\author{Mingjie Zha}
\affiliation{The Key Laboratory of Weak Light Nonlinear Photonics (Ministry of Education), School of Physics and Teda Applied Physics Institute, Nankai University, Tianjin 300071, China}
\author{Xiao Li}
\email{Corresponding author: xiao.li@cityu.edu.hk}
\affiliation{Department of Physics, City University of Hong Kong, Kowloon, Hong Kong SAR, China}
\author{Zhibo Liu}
\email{Corresponding author: zbliu@nankai.edu.cn}
\affiliation{The Key Laboratory of Weak Light Nonlinear Photonics (Ministry of Education), School of Physics and Teda Applied Physics Institute, Nankai University, Tianjin 300071, China}
\affiliation{The Collaborative Innovation Center of Extreme Optics, Shanxi University, Taiyuan, Shanxi 030006, China}

\date{\today}%
\begin{abstract}
	Rydberg excitons in transition metal dichalcogenides (TMDs) have emerged as a promising platform for investigating the properties of open quantum systems, thanks to their large binding energies (hundreds of meV).
	However, the study of Rydberg excitons in TMDs has been hindered by sample quality limitations, strong background signals from ground excitons, and broadening at room temperature.
	In this work, we report the first observation of multiple Rydberg exciton states in monolayer MoS$_2$ at room temperature using Imbert-Fedorov (IF) shift spectroscopy.
	By numerically solving the \schro equation, we extracted the quasiparticle band gaps for A and B excitons, confirming the temperature-induced redshift of the band gap, in excellent agreement with previous results.
	Our findings establish IF shift spectroscopy as a powerful tool for characterizing Rydberg excitons in TMDs, paving the way for potential applications in quantum manipulation and control.
\end{abstract}
\maketitle




\lettersection{Introduction}
Excitons in transition metal dichalcogenides (TMDs) are quasiparticles formed by electrons and holes that are tightly bound by attractive Coulomb interaction.
This strong interaction significantly enhances the light-matter interaction at the corresponding resonant energies~\cite{wuExcitonBandStructure2015,ridolfiExcitonicStructureOptical2018}.
Due to the strong spin-orbital coupling effect in TMD materials~\cite{mattheissBandStructuresTransitionMetalDichalcogenide1973,coehoornElectronicStructureMoSe1987}, the splitting of valence bands results in two types of ground excitons, commonly referred to as A and B excitons.
Analogous to Rydberg atoms, excitons possess excited states known as Rydberg excitons~\cite{biswasRydbergExcitonsTrions2023,chernikovExcitonBindingEnergy2014,hillObservationExcitonicRydberg2015}.
These Rydberg excitons are characterized by larger Bohr radii and more extensive wavefunctions compared to their ground state counterparts.
Rydberg excitons up to $n=25$ have been extensively studied in bulk copper oxide (Cu$_2$O), where they exhibit high sensitivity to external fields~\cite{zielinska-raczynskaElectroopticalPropertiesRydberg2016,heckotterHighresolutionStudyYellow2017} and charged impurities~\cite{bergenLargeScalePurification2023}.
In contrast to excitons in Cu$_2$O, the binding energies of excitons in TMD monolayers are significantly larger due to quantum confinement effects.
This unique property makes TMD monolayers ideal candidates for studying Rydberg excitons, as they offer a more extensive spectrum and enhanced visibility even at room temperature.

\begin{figure*}[t]
	\centering
	\includegraphics[width=\linewidth]{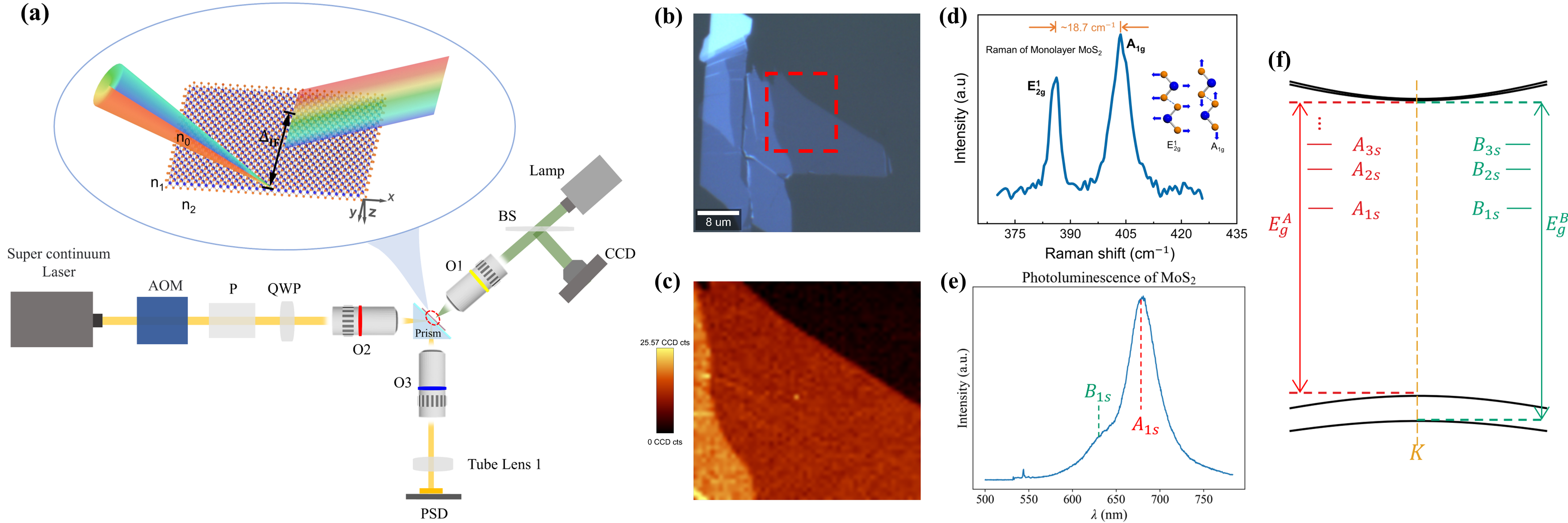}
	\caption{(a) Schematic diagram for experimental apparatus and Imbert-Fedorov shift. AOM, acousto-optic modulator; P, Glan-Taylor prism; QWP, quarter-wave plate; O1, O2, O3, three objective lenses; BS, beam splitter; PSD, position-sensitive detector; CCD, Charge-coupled Device. The inset shows the Imbert-Fedorov shift for different light wavelengths when the light passes through $n_1$ on the interface between $n_0$ and $n_2$. Results of the optical image (b), Raman map (c), Raman spectroscopy (d), and photoluminescence spectroscopy (f) of monolayer MoS$_2$ are presented, where the red dash rectangle in (b) labels the region in (c). In addition, $E_{2g}$ and $A_{1g}$ in (d) denote the in-plane and out-of-plane vibration modes. The ground excitons $A_{1s}$ and $B_{1s}$ are labeled in (f). (e) the schematic diagram of Rydberg excitons in MoS$_2$. Rydberg states and the band gap are clearly shown.
	}
	\label{fig:experiment-apparatus}
\end{figure*}

There have been many investigations of Rydberg excitons in TMD materials, including MoS$_2$, WS$_2$, MoSe$_2$, MoTe$_2$, and WSe$_2$, which are mainly based on various optical measurements such as photoluminescence spectroscopy~\cite{hillObservationExcitonicRydberg2015,pandeyUnravelingBiexcitonExcitonic2019}, differential reflectivity spectroscopy~\cite{robertOpticalSpectroscopyExcited2018,shreeHighOpticalQuality2019}, and photocurrent spectroscopy~\cite{vaqueroExcitonsTrionsRydberg2020}.
The increased binding energies and the potential for room-temperature observation of Rydberg excitons in TMD monolayers open up new possibilities for investigating their properties and exploring their potential applications in various optoelectronic devices and quantum technologies.
For exampmle, benefiting from the sensitivity of Rydberg excitons to the surrounding dielectric environment and quasiparticle interactions,
there emerged new investigations that are focused on utilizing A$_{2s}$ excitons in WSe$_2$ and WS$_2$ to characterize the correlated insulating states in moir\'e superlattices,
including twisted bilayer graphene~\cite{huObservationRydbergMoire2023,guRemoteImprintingMoire2024,heDynamicallyTunableMoire2024}.
The Rydberg excitons exhibit clear responses to the correlated insulating states in neighboring moir\'e superlattices in these observations.
This indicates that Rydberg excitons can be a potential optical sensor to detect quantum states in the adjacent materials and thus can work as a probe to study the electronic properties of the quantum system.

However, the observation of even higher Rydberg exciton levels in TMD monolayers at room temperature is challenging.
For example, higher Rydberg excitonic states such as A$_{3s}$ or B$_{2s}$ states are usually ambiguous in the optical spectrum due to strong background resonances arising from ground excitons or dissipation through exciton-phonon channels under room temperature.
Consequently, a radically new spectroscopic technique is required to effectively probe the signatures of multiple higher Rydberg excitons, regardless of temperature or ground state signals.
In this work, we report the observation of multiple Rydberg excitons at room temperature in monolayer MoS$_2$ by employing the Imbert-Fedorov (IF) shift spectroscopy.
Through theoretical calculations based on thin-film and conductivity models, we managed to fit the peaks in the IF spectrum by multiple Lorentzian peaks, giving out the conductivity of monolayer MoS$_2$.
Excitonic signatures in the conductivity are then modeled by solving the \schro equation with Rytova-Keldysh potential, which confirms the excitons up to B$_{3s}$ state.
These observations reveal that the Rydberg excitons in TMD monolayers can be effectively probed by the IF shift spectroscopy, which opens up new possibilities for studying the properties of Rydberg excitons in TMD monolayers.



\lettersection{Experimental setup}
Incident light with a finite waist experiences spatial displacements through the interface between two media. The transverse part of the spatial displacements is called Imbert-Fedorov shift (IF shift)~\cite{onodaHallEffectLight2004,aielloRoleBeamPropagation2008,bliokhGoosHanchenImbert2013}, which is related to the conversion between spin and orbital angular momentum of the light, referred to as photonic spin Hall effect (PSHE)~\cite{liuPhotonicSpinHall2022}.
However, the IF shifts are typically of sub-wavelength scale, making it difficult to measure using traditional methods.
As a result, the first observation of spin-dependent IF shifts in PSHE was successfully achieved by Hosten and Kwiat using quantum weak measurements~\cite{hostenObservationSpinHall2008}.
However, quantum weak measurement cannot provide more information about the optical properties of the materials beyond single-wavelength responses, which hinders the application of the IF shifts in the spectroscopic investigation.
Very recently, a new characterization method equipped with a super-continuum light source was developed with the usage of beam displacement amplification technique (BDAT), which achieved the switch of different light wavelengths~\cite{zhaOpticalShiftSpectroscopy2024}.
The spectral measurement of IF shifts in two-dimensional materials can be achieved with a supercontinuum light source and the BDAT at room temperature (Fig.~\ref{fig:experiment-apparatus}(a)). We placed monolayer MoS$_2$ at the interface between BK-7 glass and air. After passing through the acousto-optic modulator (AOM), the light is polarized into left- or right-circularly polarized states by the quarter-wave plate (QWP). Then, the incident light is focused on the monolayer MoS$_2$. The reflected light is collected by an objective lens (O3) and recorded by a position-sensitive detector (PSD).

The optical image, and the Raman map of the sample (of the dashed red zone) are displayed in Fig.~\ref{fig:experiment-apparatus}(b) and Fig.~\ref{fig:experiment-apparatus}(c), respectively. Raman spectroscopy is conducted on monolayer MoS$_2$ samples (Fig.~\ref{fig:experiment-apparatus} (d)), and the Raman signatures confirmed the samples to be monolayer~\cite{zengValleyPolarizationMoS22012}. We scanned the samples to obtain the Raman map at the mode $A_{1g}$, and we found the sample to be uniform and clean, which precludes the impacts of the impurities or defects (Fig.~\ref{fig:experiment-apparatus}(c)).
We observed the prominent peak A and the shoulder peak B by photoluminescence spectroscopy (Fig.~\ref{fig:experiment-apparatus} (f)), which arise from $A_{1s}$ and $B_{1s}$ excitons, respectively. These observations agree with the results in other experiments~\cite{makAtomicallyThinMathrm2010,splendianiEmergingPhotoluminescenceMonolayer2010}. Like Rydberg atoms, ground excitonic states have their excited states with the principal quantum number $n>1$, which will also impact the optical properties of the 2D TMDs. However, due to the strong background signals from ground excitons, Rydberg excitons are usually hidden under the mask in optical absorption or photoluminescence experiments.
With the sensitivity of the IF shift to the optical properties of the interface, it is feasible to study how the signatures of Rydberg excitons in 2D TMDs manifest in IF shift measurements.

\begin{figure}[t]
	\centering
	\includegraphics[width=\linewidth]{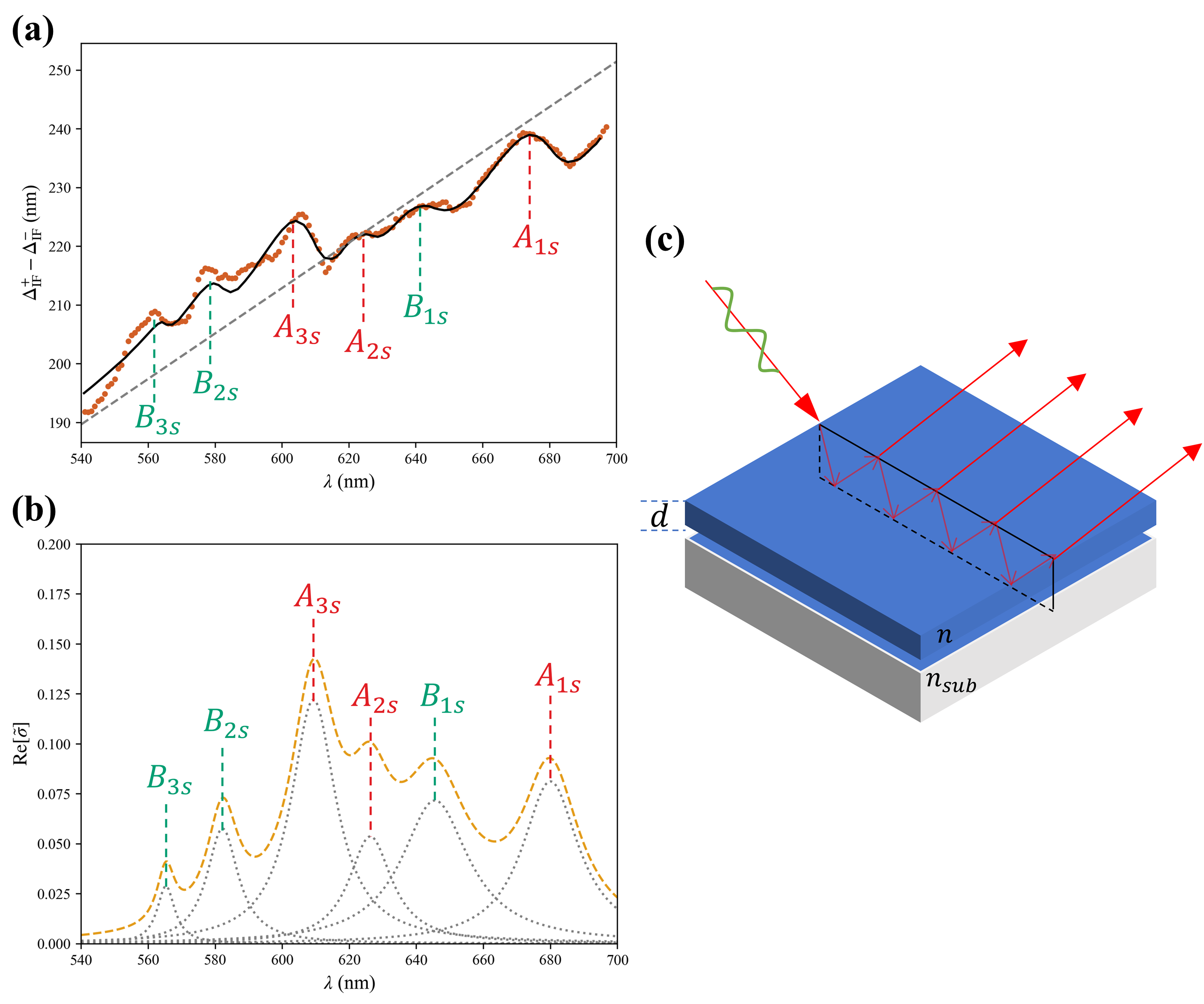}
	\caption{(a) Experimental IF shift (orange dots), theoretical IF shift from the Lorentzian peaks in permittivity (black line) and from the single-particle contributions from monolayer MoS$_2$ (grey dashed line). Six peaks are clearly exhibited in experiment results and labeled by corresponding Rydberg excitonic states. (b) reduced conductivity $\tilde{\sigma}$ based on the permittivity comprising of six fitted Lorentzian peaks, which correspond to Rydberg excitons' labels. (c) Schematic diagram of the thin-film model. The incident light experiences multiple reflections in the material with thickness d and refractive index n in (a).
	}
	\label{fig:theory-models}
\end{figure}

We now present the result of the IF shift spectroscopy measurement in monolayer MoS$_2$ in Fig.~\ref{fig:theory-models}(b).
Six prominent peaks are visible around \SI{670}{nm}, \SI{640}{nm}, \SI{620}{nm}, \SI{605}{nm}, \SI{580}{nm}, and \SI{560}{nm}.
In contrast, the IF shift spectroscopy in a bulk MoS$_2$ sample showed only two peaks at \SI{580}{nm} and \SI{650}{nm}, respectively~\cite{suppinfo}, confirming the signature of six peaks is exclusive to the monolayer sample.
In the following, we will use theoretical modeling to study whether these six peaks could be attributed to Ryderberg excitons in the sample.

\lettersection{Theoretical modeling}
The IF shift can be calculated theoretically by the following formula~\cite{hermosaReflectionBeamshiftsVisible2016}
\begin{equation}
	\begin{split}
		\Delta^{IF} = -\frac{\cot\theta_0}{k_0} \Big[
		2\sqrt{w^s w^p}\sin(\eta-\phi^p + \phi^s) + \\ \frac{
				w^p (a^s)^2 + w^s (a^p)^2
			}{
				a^p a^s
			} \sin\eta
			\Big],
	\end{split}
	\label{eq:IF-shift}
\end{equation}
where $\theta_0$ is the incident angle, $k_0=2\pi/\lambda$ is the wave vector of the incident light, and $s$ and $p$ label the out-of-plane and the in-plane directions with respect to the incident plane.
Moreover, $a^{s,p}$ is the amplitude of electric field along $s$- and $p$-direction while $\eta$ is the phase difference between the two polarizations.
For circularly polarized light, $a^s=a^p=1$ and $\eta=\pi/2$ ($\eta = -\pi/2$) for left (right) circularly polarized light.
Finally, $w^{s,p}=\frac{(R^{s,p})^2 (a^{s,p})^2}{(R^s)^2 (a^s)^2 + (R^p)^2 (a^p)^2}$, where $r^{s,p}=R^{s,p}e^{i \phi^{s,p}}$ is the complex Fresnel coefficient of $s$- and $p$-polarized light components.

Based on Eq.~\eqref{eq:IF-shift}, the IF shift will be dominated by two quantities.
First, the $k_0$ of the incident light will provide an overall linear dependence on the wavelength $\lambda$.
Second, the IF shift is controlled by the Fresnel coefficients of the material, which characterizes the material's optical responses to the incident light.
There are two models for calculating the Fresnel coefficients: the thin-film model~\cite{heavensOpticalPropertiesThin1960,liAbnormalSpatialShifts2021} and the conductivity model~\cite{groscheGoosHanchenImbertFedorovShifts2015,zhanTransferMatrixMethod2013}.
In the thin-film model, the sample is modeled with a finite thickness (Fig.~\ref{fig:theory-models} (a)).
In contrast, the sample is treated as an infinitely thin conducting layer at the interface in the conductivity model.
In the following, we presented the theoretical calculations based on the thin-film model, while the results of the conductivity model and the comparison between the two models can be referred to in Ref.~\cite{suppinfo}.

The thin-film model treats the material as a layer with thickness $d$, within which we can calculate the Fresnel coefficients $r^{s,p}$ by thin-film interference or transfer-matrix method.
After some algebra, we can obtain the complex Fresnel coefficients $r^{s,p}$ of the whole structure as
\begin{equation}
	r_{s,p} = \frac{
		r_{01}^{s,p} + r_{12}^{s,p} e^{i2\delta}
	}{
		1 + r_{01}^{s,p} r_{12}^{s,p} e^{i2\delta}
	},
	\label{eq:thin-film-rsp}
\end{equation}
where $r_{jk}^s = \frac{n_j \cos \theta_j - n_k \cos \theta_k}{n_j \cos \theta_j + n_k \cos \theta_k}$,
$r_{jk}^s = \frac{n_k \cos \theta_j - n_j \cos \theta_k}{n_k \cos \theta_j + n_j \cos \theta_k}$ are the Fresnel coefficients at the interface between media $j$ and $k$.
Moreover, $\theta_{j,k}$ is the incident angle in medium $j$ and $k$, while $n_{j,k}$ is the complex refractive indexes of medium $j$ and $k$, which is related to the permittivity $n_{j,k} = \sqrt{\varepsilon_{j,k}}$.
Finally, $\delta=2\pi n_1 d \cos \theta_1/\lambda$.
Once we know about the permittivity $\varepsilon_{0,1,2}$, we can directly calculate the overall Fresnel coefficients based on Eq.~\eqref{eq:thin-film-rsp}. In this experiment, index 0, 1, 2 refer to the substrate BK-7 glass, the monolayer MoS$_2$ and the air, respectively.

Now, the key step to calculate the IF shift reduces to the evaluation of the permittivity $\varepsilon$, defined as
\begin{equation}
	\varepsilon = \varepsilon_\infty + \frac{
		i \tilde{\sigma} c
	}{
		d\omega
	},
	\label{epsilon-to-cond}
\end{equation}
where $d=\SI{0.65}{nm}$ is the thickness of the monolayer MoS$_2$, and $c$ is the speed of light.
Moreover, $\varepsilon_\infty$ is the background permittivity in MoS$_2$, which is taken to be $\varepsilon_\infty$=15.6~\cite{asadiStructuralElectronicMechanical2019}.
Finally, $\tilde{\sigma}$ is the reduced conductivity.

As a sanity check, we calculate the IF shift arising from the single-particle band structure of monolayer MoS$_2$, which is shown as the grey dashed line in Fig.~\ref{fig:theory-models}.
This is achieved by replacing $\tilde{\sigma}$ in the permittivity with the optical conductivity arising from the single-particle band structure of monolayer MoS$_2$~\cite{wuExcitonBandStructure2015}.
We thus find that the contribution to IF from the band structure of monolayer MoS$_2$ is featureless, only proportional to $k_0^{-1}$.
Therefore, we suspect that the peaks in the experimental data could arise from the Rydberg excitons in monolayer MoS$_2$.


The IF shift spectra we obtained provide abundant excitonic information of monolayer MoS$_2$ by presenting multiple prominent peaks that differ from traditional optical measurements such as absorption, photoluminescence, etc. To extract the information about the excitons of monolayer MoS$_2$, we used multiple Lorentzian peaks to expand the permittivity to fit the IF shift spectrum peaks~\cite{liMeasurementOpticalDielectric}
\begin{equation}
	\varepsilon = \varepsilon_\infty + \sum_i \frac{
		\omega_{a_i}^2
	}{
		\omega_{c_i}^2 - \omega^2 - i\gamma_i \omega
	},
	\label{eq:epsilon-expansion}
\end{equation}
where $\omega_{a_i}$, $\omega_{c_i}$, $\gamma$ are the amplitude, center, and broadening of $i$-th Lorentzian peak, $\omega$ is the frequency of the incident light.
The Lorentzian peaks are set to be around the observed IF shift peaks, with adjustable Lorentzian parameters $\omega_{a_i}$, $\omega_{c_i}$, $\gamma$.
Then, the nonlinear fit algorithm is used to fit the experiment results by tuning the three sets of parameters in the thin-film model.
The fitted parameters are shown in Table~\ref{Table}.
The fitted Lorentzian peaks capture the main peak signatures in the IF spectrum very well, as shown in Fig.~\ref{fig:theory-models}(b).
Finally, we convert the fitted permittivity results are converted to reduced conductivity $\tilde{\sigma}$, which are shown as yellow dashed lines in Fig.~\ref{fig:theory-models}(b).

\begin{table}[]
	\begin{tabular}{c|c|c|c}
		\hline\hline
		         & $\omega_{c_i}$ (meV) & $\omega_{a_i}$ (meV) & $\gamma_i$ (meV) \\ \hline
		$A_{1s}$ & 1823.54              & 1167.14              & 55.20            \\ \hline
		$B_{1s}$ & 1920.86              & 1258.77              & 72.76            \\ \hline
		$A_{2s}$ & 1980.00              & 898.06               & 49.42            \\ \hline
		$A_{3s}$ & 2035.15              & 1420.83              & 54.42            \\ \hline
		$B_{2s}$ & 2130.00              & 856.38               & 41.85            \\ \hline
		$B_{3s}$ & 2193.62              & 468.36               & 25.00            \\
		\hline\hline
	\end{tabular}
	\caption{\label{Table}
		The fitting of the thin film model to the experimental IF shift spectra using Eq.~\eqref{eq:epsilon-expansion}. }
\end{table}

We now argue that these Lorentzian peaks are associated with the Rydberg excitons in monolayer MoS$_2$, which can be classified into the A and B series, as labeled in Fig.~\ref{fig:theory-models}(b).
Up to $n=3$ states for both A and B series can be clearly observed in our experiment under room temperature, which are different from past results obtained by permittivity~\cite{ermolaevBroadbandOpticalProperties2020,liBroadbandOpticalProperties2014,morozovOpticalConstantsDynamic2015,diwareCharacterizationWaferscaleMoS2017}, refractive indexes~\cite{zhangMeasuringRefractiveIndex2015, islamInPlaneOutofPlaneOptical2021}, or conductivity~\cite{songComplexOpticalConductivity2019,jiaExcitonicQuantumConfinement2016} measurements.
Most of the experiments are simply showing A and B excitonic signatures.
Several show Rydberg excitons under low temperatures.
Our IF shift spectrum unprecedentedly reveals the A and B Rydberg excitons under room temperature.



\begin{figure}[t]
	\centering
	\includegraphics[width=0.9\linewidth]{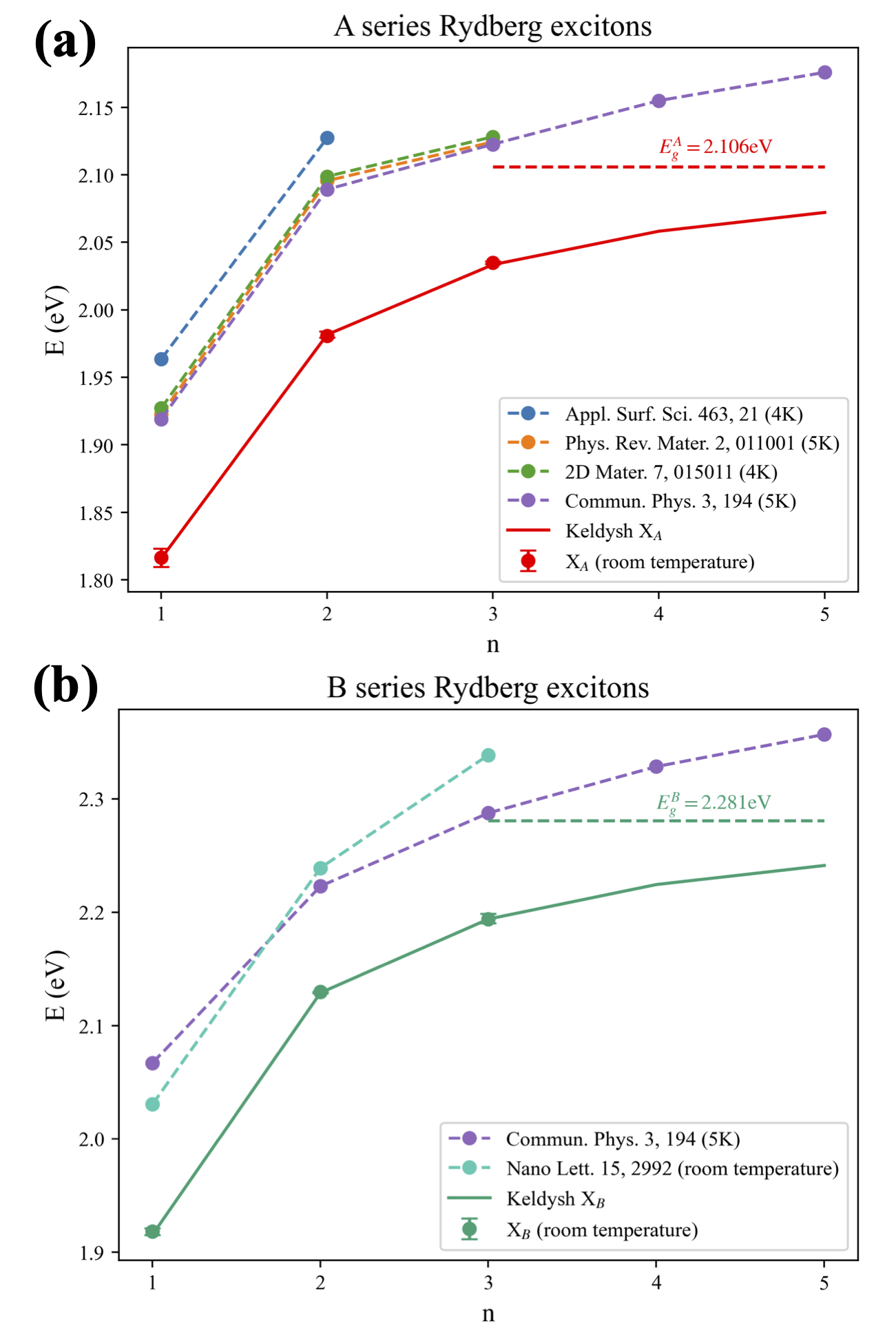}
	\caption{A (a) and B (b) series Rydberg excitons from our experiments (X$_A$, X$_B$) and other references. The variations of center energies of Lorentzian peaks across samples are expressed as error bars, which are negligible. Based on the \schro equation integrated with Rytova-Keldysh interaction potential, we fitted the energies of the Lorentzian peaks and plotted the fitted result Keldysh X$_{A,B}$ up to n=5. From the fitting, we also obtained the quasiparticle band gap for the A and B series, as shown in (a) and (b).}
	\label{fig:rydberg-excitons-compare}
\end{figure}

\lettersection{Extracting the properties of the Rydberg excitons}
The extracted exciton energies in Table~\ref{Table} allow us to further elucidate the properties of the Rydberg excitons, such as the band gap $E_g^{A,B}$ and the screening length $r_0^{A,B}$ for the A and B excitons.
This can be done by fitting these exciton energies to the two-dimensional \schro equation with the Rytova-Keldysh potential, which is known to yield Rydberg excitons~\cite{chernikovExcitonBindingEnergy2014,gorycaRevealingExcitonMasses2019}.
The radial part of the equation reads as
\begin{equation}
	H \psi_{ns} (r) = (E_n - E_g) \psi _{ns} (r),
	\label{eq:schro-eigen}
\end{equation}
where $E_g$ is the band gap and $\psi_{ns}$ is the wavefunction of the exciton.
The Hamiltonian $H$ contains the kinetic energy and the interaction potential $V(r)$
\begin{equation}
	H = - \frac{\hbar^2}{2\mu} \dfrac{d^2}{dr^2} +V(r),
	\label{eq:hamiltonian}
\end{equation}
where $\mu$ is the effective mass of the exciton.
We take $\mu_A=0.25\mu_0$
and $\mu_B=0.28\mu_0$~\cite{berkelbachTheoryNeutralCharged2013}, where $\mu_0$ is the free electron mass.
The Rytova-Keldysh potential~\cite{keldyshCoulombInteractionThin1979b,rytovaScreenedPotentialPoint1967} reads as
\begin{equation}
	V(r) = - \frac{
		\pi e^2
	}{
		2 r_0
	} \left[
		\mathcal{H}_0 \left(\frac{\kappa r}{r_0}\right) - \mathcal{Y}_0  \left(\frac{\kappa r}{r_0}\right)
		\right],
	\label{eq:keldysh-rytova-potential}
\end{equation}
where $\mathcal{H}_0$ is the zeroth-order Struve function while $\mathcal{Y}_0$ is the zeroth-order Bessel function of the second kind.
Moreover, $r_0$ is the screening length, which is a key fitting parameter.
Finally, $\kappa=(\varepsilon_1 + \varepsilon_2)/2$ is the average dielectric constant of the media encapsulating the material~\cite{chernikovExcitonBindingEnergy2014}.
In the present paper, we take $\varepsilon_1=2.3$ for the BK-7 glass over optical frequency while $\varepsilon_2=1$ for air. This consideration fixes $\kappa=1.65$.

The fitted A and B series Rydberg excitons are shown in Fig.~\ref{fig:rydberg-excitons-compare}.
The error bars of both A and B Rydberg excitons are small, showing good consistency across different samples.
We first discuss the two fitted band gaps $E_g$, which is  $E_g^A=\SI{2.106}{eV}$ for the A Rydberg exciton and $E_g^B=\SI{2.281}{eV}$ for the B Rydberg exciton. 
Assuming the conduction band splitting to be $\Delta E_{CB} \approx \SI{15}{meV}$~\cite{robertMeasurementSpinforbiddenDark2020,pisoniInteractionsMagnetotransportSpinValley2018}, we can deduce the valence band splitting as $\Delta E_{VB}=E_g^B-E_g^A-\Delta E_{CB}=\SI{156}{meV}$, consistent with the theoretical calculations~\cite{miwaElectronicStructureEpitaxial2015,ramasubramaniamLargeExcitonicEffects2012}.
From these results, we estimate that the binding energy for ground A exciton is $\sim\SI{290}{meV}$, and that for ground B exciton is $\sim\SI{360}{meV}$ in our sample.
In Fig.~\ref{fig:rydberg-excitons-compare}, we compare our results with those in other references.
We observe an overall red shift in our results, which is likely due to temperature-dependent band gap renormalization~\cite{varshniTemperatureDependenceEnergy1967,zhangAbsorptionLightExcitons2014}.
Next, we discuss the optimal screening lengths, which are found to be $r_0^A=\SI{7.01}{nm}$ and $r_0^B=\SI{5.30}{nm}$.
The large screening lengths arise from the dielectric environment of the excitons in our samples (BK-7 and air), which is different from samples encapsulated by h-BN.
Similar screening lengths are also observed in other experiments involving WS$_2$ samples encapsulated by SiO$_2$~\cite{chernikovExcitonBindingEnergy2014}.

\lettersection{Discussion}
In summary, we applied Imbert-Fedorov shift spectroscopy to observe both the A and B series Rydberg excitons at room temperature for the first time.
By utilizing thin-film and conductivity models and expanding the permittivity or conductivity with Lorentzian peaks, we successfully fitted the experimental data for the IF shift.
Furthermore, we numerically solved the \schro equation with the Keldysh interaction potential to extract exciton parameters from the data.
The obtained band gaps and screening lengths for A and B series excitons align with other experimental observations, taking into account the temperature factor.

The observations of Rydberg excitons under room temperature are plausible and feasible, because the energy separations between different exciton levels generally exceed $\SI{25}{meV}$.
However, the simultaneous observation of both A and B Rydberg excitons under these conditions is an exceptional result, rarely reported in the literature.
Our experiment not only revealed both A and B Rydberg excitons up to $3s$ states at room temperature but also demonstrated the sensitivity of IF shift spectroscopy to the optical response of materials, making it an ideal tool for detecting and characterizing their fine structures.
Unlike traditional optical measurements such as absorption, photoluminescence, and differential reflectivity, which often struggle with the broadening of ground and Rydberg excitons, IF spectroscopy stands out in identifying weak signals hidden behind strong backgrounds.

There are many possibilities for further improvements to this groundbreaking technique.
By designing a cooling chamber for low-temperature experiments, applying gate voltages to adjust carrier density, or introducing external factors such as magnetic or electric fields, we can push the boundaries of our understanding even further.
Under these conditions, higher-level Rydberg excitons will persist at low temperatures due to suppressed dissipation through electron-phonon interaction, while trions' responses can be captured by IF spectroscopy under different doping densities.
Moreover, by designing various structures, we can study the quantum states emerging in moir\'{e} superlattices using Rydberg excitons or trions under doping, as demonstrated by recent studies~\cite{huObservationRydbergMoire2023,guRemoteImprintingMoire2024,heDynamicallyTunableMoire2024}.
The future of IF spectroscopy is highly promising, as it has the potential to become a sensitive characterization method for detecting quasiparticle interactions in quantum systems, paving the way for groundbreaking discoveries in the field of condensed matter physics.

\lettersection{Acknowledgement}
This work was supported by the Natural Science Foundation of China (Grant 12174207). X.L. is supported by the Research Grants Council of Hong Kong (Grants No.~CityU 11300421, CityU 11304823, and C7012-21G) and City University of Hong Kong (Project No. 9610428).

\bibliography{apssamp}

\end{document}